\documentclass{emulateapj}
\usepackage{apjfonts}
\lefthead{PARK ET AL.} 
\righthead{LENSING SENSITIVITY TO HABITABLE-ZONE PLANETS}

\newcommand{\re}{r_{\rm E}}

\begin{document}
\title{Microlensing Sensitivity to Earth-mass Planets in the Habitable Zone}

\author{
Byeong-Gon Park\altaffilmark{1}, 
Young-Beom Jeon\altaffilmark{1},
Chung-Uk Lee\altaffilmark{1} and 
Cheongho Han\altaffilmark{2,3}
}
\altaffiltext{1}{Korea Astronomy and Space Science Institute, Hwaam-Dong, 
Yuseong-Gu, Daejeon 305-348, Korea; bgpark,ybjeon,leecu@kasi.re.kr}
\altaffiltext{2}{Department of Physics, Institute for Basic Science
Research, Chungbuk National University, Chongju 361-763, Korea; 
cheongho@astroph.chungbuk.ac.kr}
\altaffiltext{3}{corresponding author}

% ==================================================================

%\submitted{Submitted to The Astrophysical Journal}

\begin{abstract}
Microlensing is one of the most powerful methods that can detect 
extrasolar planets and a future space-based survey with a high 
monitoring frequency is proposed to detect a large sample of 
Earth-mass planets.  In this paper, we examine the sensitivity 
of the future microlensing survey to Earth-mass planets located 
in the habitable zone. For this,  we estimate the fraction of 
Earth-mass planets that will be located in the habitable zone of 
their parent stars by carrying out detailed simulation of 
microlensing events based on standard models of the physical and 
dynamic distributions and the mass function of Galactic matter.
From this investigation, we find that among the total detectable 
Earth-mass planets from the survey, those located in the habitable 
zone would comprise less than 1\% even under a less-conservative 
definition of the habitable zone.  We find the main reason for the 
low sensitivity is that the projected star-planet separation at 
which the microlensing planet detection efficiency becomes maximum 
(lensing zone) is in most cases substantially larger than the median 
value of the habitable zone.  We find that the ratio of the median 
radius of the habitable zone to the mean radius of the lensing zone 
is roughly expressed as
$d_{\rm HZ}/r_{\rm E}\sim 0.2(m/0.5\ M_\odot)^{1/2}$.

\end{abstract}


\section{Introduction}

Since the first detection by using the pulsar timing method 
\citep{wolszczan92}, extrasolar planets have been and are going to be 
detected by using various techniques including the radial velocity 
technique \citep{mayor95, marcy96}, transit method \citep{struve52}, 
astrometric technique \citep{sozzetti05}, direct imaging \citep{angel94, 
stahl95}, and microlensing \citep{mao91, gould92}.  See the reviews of 
\citet{perryman00, perryman05}.

The microlensing signal of a planet is a short-duration perturbation 
to the smooth standard light curve of the primary-induced lensing 
event occurring on a background source star.  The planetary lensing 
signal induced by a giant planet with a mass equivalent to that of 
the Jupiter lasts for a duration of $\sim 1$ day, and the duration 
decreases in proportion to the square root of the mass of a planet, 
reaching several hours for an Earth-mass planet.  Once the signal is 
detected and analyzed, it is possible to determine the planet/primary 
mass ratio, $q$, and the projected primary-planet separation, $s$, 
in units of the Einstein ring radius $\re$, which is related to the 
mass of the lens, $m$, and distances to the lens and source, $D_L$ 
and $D_S$, by
\begin{equation}
\re \simeq 4.9\ {\rm AU} 
\left( {m\over 0.5\ M_\odot}\right)^{1/2}
\left( {D_L\over 6\ {\rm kpc}}\right)^{1/2}
\left( 1-{D_L\over D_S}\right)^{1/2}.
\label{eq1}
\end{equation} 
Currently, several experiments are going underway to search for 
planets by using the microlensing technique and two robust detections 
of Jupiter-mass planets were recently reported by \citet{bond04} and 
\citet{udalski05}.  In addition, a future space-based survey with the 
capability of continuously monitoring stars at high cadence by using 
very large format imaging cameras is proposed to detect a large sample 
of Earth-mass planets \citep{bennett02}.

The microlensing technique has various advantages over other methods.
First, microlensing is sensitive to lower-mass planets than most other
methods and it is possible, in principle, to detect Earth-mass planets 
from ground-based observations with adequate monitoring frequency and 
photometric precision \citep{gould04}.  Second, a large sample of planets, 
especially low-mass terrestrial planets, will be detected at high S/N 
with the implementation of future lensing surveys, and thus the 
microlensing technique will be able to provide the best statistics of 
Galactic population of planets.  Third, the planetary lensing signal 
can be produced by the planet itself, and thus the microlensing 
technique is the only proposed method that can detect and characterize 
free-floating planets \citep{bennett02, han05a}.  Fourth, the microlensing 
technique is distinguished from other techniques in the sense that the 
planets to which it is sensitive are much more distant than those found 
with other techniques and the method can be extended to search for planets 
located even in other galaxies \citep{covone00, baltz01}.

In addition to the detections of planets, the habitability of the 
detected planets is of great interest. One basic condition for the 
habitability is that  planets should be located at a suitable distance 
from their host stars.  Then, a question is whether a significant 
fraction of planets to be detected by future lensing surveys would be
in the lensing zone.  If the planets in the habitable zone comprise 
a significant fraction of the total microlensing sample of planets, 
a statistical analysis on the frequency of terrestrial planets in the 
habitable zone would be possible under some sorts of assumptions about 
the shape and inclination of the planetary orbit around host stars.  
In this paper, we examine the sensitivity of future lensing surveys 
to Earth-mass planets located in the habitable zone by carrying out 
detailed simulation of Galactic microlensing events based on the 
standard models of the physical and dynamic distributions and the mass 
function of Galactic matter.  The sensitivities to habitable-zone 
planets of other planet search methods were discussed by \citet{gould03b} 
(transit method), \citet{gould03a} (astrometric method),  and 
\citet{sozzetti02} (space interferometry).

The paper is organized as follows.  In \S\ 2, we briefly describe the 
basics of planetary microlensing.  In \S\ 3, we describe the procedure 
of the simulation of Galactic microlensing events produced by lenses 
with Earth-mass planet companions.  We then describe the procedure of 
estimating the fraction of planets located in the habitable zone among 
the total number of detectable Earth-mass planets based on the planetary 
lensing events produced by the simulation.  In \S\ 4, we present the 
results of the simulation and discuss about the results.  We conclude 
in \S\ 5.

\section{Basics of Planetary Microlensing}

Planetary lensing is an extreme case of binary lensing with a very 
low-mass companion.  Because of the very small mass ratio, planetary 
lensing behavior is well described by that of a single lens of the 
primary for most of the event duration.  However, a short-duration 
perturbation can occur when the source star passes the region around 
caustics.

The caustic is an important feature of binary lensing and it represents 
the source position at which the lensing magnification of a point 
source becomes infinite.  The caustics of binary lensing form a single 
or multiple closed figures where each of which is composed of concave 
curves (fold caustics) that meet at cusps.  For a planetary case, 
there exist two sets of disconnected caustics.  One small `central' 
caustic is located close to the primary lens, while the other bigger 
`planetary' caustic(s) is (are) located away from the primary at 
the position with a separation vector from the primary lens of 
\begin{equation}
{\bf r}={\bf s}\left(1-{1\over s}\right)^2,
\label{eq2}
\end{equation} 
where ${\bf s}$ is the position vector of the planet from the primary
lens.  Then, the caustics are located within the Einstein ring when 
the planet is located within the separation range of $0.6\lesssim s
\lesssim 1.6$, which is often referred as the `lensing zone'.  The 
number of the planetary caustics is one or two depending on whether 
the planet lies outside ($s>1$) or inside ($s<1$) the Einstein ring.

The size of the caustic, which is directly proportional to the planet 
detection efficiency, is dependent on both $q$ and $s$.  Under 
perturbative approximation (when $q\ll 1$ and $|s-1|\gg q$), the sizes 
of the central ($\Delta r_{\rm cc}$) and planetary ($\Delta r_{\rm pc}$) 
caustics as measured by the width along the star-planet axis are 
represented, respectively, by
\begin{equation}
{\Delta r_{\rm cc}\over \re} \simeq {4q\over (s-s^{-1})^2} \rightarrow 
\cases{
4 q s^{-2}, & for $s>1$, \cr
4 q s^2,    & for $s<1$, \cr
}
\label{eq3}
\end{equation}
and 
\begin{equation}
{\Delta r_{\rm pc}\over \re} \simeq 
\cases{
4q^{1/2}[s(s^2-1)^{1/2}]^{-1} \rightarrow 4 q^{1/2}s^{-2}, 
& for $s>1$, \cr
2q^{1/2}(\kappa_0-\kappa_0^{-1}+\kappa_0s^{-2})\cos\theta_0
\rightarrow 1.3 q^{1/2}s^3, & for $s<1$, \cr
}
\label{eq4}
\end{equation}
where 
$\kappa(\theta)=\{[\cos 2\theta\pm(s^4-\sin^2 2\theta)^{1/2}]/(s^2-
1/s^2)\}^{1/2}$, $\theta_0=[\pi \pm \sin^{-1}({\sqrt{3}s^2/2})]/2$, 
and $\kappa_0=\kappa(\theta_0)$ \citep{bozza00, an05, chung05, han06}.  
The size of the caustic becomes maximum when $s\sim 1$ and decreases 
rapidly as the separation becomes bigger ($\propto s^{-2}$ for both 
the central and planetary caustics) or smaller ($\propto s^2$ for the 
central caustic and $\propto s^3$ for the planetary caustic) than the 
Einstein ring radius.  As a result, only planets located within the 
lensing zone have non-negligible chance of producing planetary signals.

\begin{figure}[t]
\epsscale{1.2}
\plotone{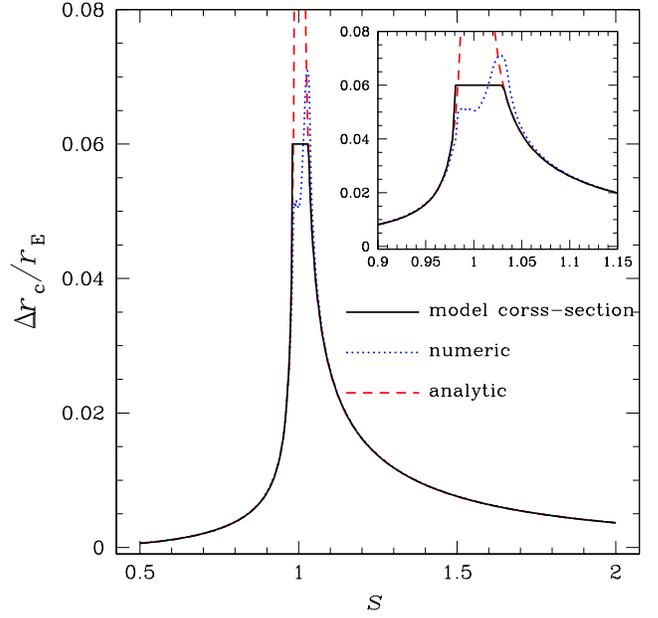}
\caption{\label{fig:one}
The variation of the  
caustic size as a function of the primary-planet separation, $s$, for 
a planetary lens system with a mass ratio of $q=10^{-5}$.  The dotted 
curve is calculated numerically while the dashed curve is computed by 
using the analytic formalism based on perturbative approximation 
(Eqs.~[\ref{eq3}] and [\ref{eq4}]).  The two results match very well 
except the region around $s=1$, where the perturbative approximation is 
no longer valid.  The solid curve is the model adopted in our simulation 
as the relative planet detection efficiency as a function of $s$.
}\end{figure}

In Figure~\ref{fig:one}, we present the variation of the total caustic 
size, $\Delta r_{\rm c} =\Delta r_{\rm cc}+\Delta r_{\rm pc}$, as a 
function of the primary-planet separation for a planetary lens system 
with a mass ratio of $q=10^{-5}$, which corresponds to an Earth-mass 
planet around a star with $0.3\ M_\odot$.  In the figure, the dotted 
curve is calculated numerically while the dashed curve is computed  
analytically by using the formalism in equations~(\ref{eq3}) and 
(\ref{eq4}).  The disagreement between the two curves in the region 
around $s=1$ is due to the failure of the perturbative approximation 
in this region.

% Figure 1 --------------------------------------------------------------

\section{Simulation}

To estimate the fraction of planets located in the habitable zone 
among the total number of Earth-mass planets detectable by future 
lensing surveys, we conduct detailed simulation of Galactic 
microlensing events.  In the simulation, we assume that the survey 
is conducted toward the Baade's Window field centered at the Galactic 
coordinates of $(l,b)=(1^\circ,-4^\circ)$.  The absolute brightnesses 
of the source stars are assigned on the basis of the luminosity function 
of \citet{holtzman98} constructed by using the {\it Hubble Space 
Telescope}.  Once the absolute magnitude is assigned, the apparent 
magnitude is determined considering the distance to the source star 
and extinction.  The extinction is determined such that the source 
star flux decreases exponentially with the increase of the dust column 
density.  The dust column density is computed on the basis of an 
exponential dust distribution model with a scale height of $h_z=120\ 
{\rm pc}$, i.e.\ $\propto \exp(-|z|/h_z)$, where $z$ is the distance 
from the Galactic plane.  We normalize the amount of extinction so 
that $A_V=1.28$ for a star located at $D_{\rm S}=8\ {\rm kpc}$ 
following the measurement of \citet{holtzman98}.  The proposed 
space-based lensing survey searching for Earth-mass planets will 
monitor main-sequence stars to minimize the finite-source effect, 
which washes out the planetary lensing signal.  We, therefore, assume 
that the brightness range of source stars to be monitored by the 
survey is $18.5\lesssim V\lesssim 25.0$, which corresponds to early 
F to early M-type main-sequence stars.

The locations of the source stars and lens matter are allocated based 
on the standard mass distribution model of \citet{han03}. In the model, 
the bulge mass distribution is scaled by the deprojected infrared 
light density profile of \citet{dwek95}, specifically model G2 with 
$R_{\rm max}=5$ kpc from their Table 2.  The velocity distribution 
of the bulge is deduced from the tensor virial theorem and the 
resulting distribution of the lens-source transverse velocity is 
listed in Table 1 of \citet{han95}, specifically non-rotating barred 
bulge model.  The disk matter distribution is modeled by a 
double-exponential law, which is expressed as $\rho(R,z) =\rho_0 
\exp [-(r-R_0)/h_R+|z|/h_z]$, where $(R,z)$ is the Galactocentric 
cylindrical coordinates, $R_0=8$ kpc is the distance of the sun from 
the Galactic center, $\rho_0=0.06\ M_\odot\ {\rm pc}^{-3}$ is the mass 
density in the solar neighborhood, and $h_R=3.5$ kpc and $h_z=325$ pc 
are the radial and vertical scale heights.  According to the models 
of the dynamical and physical distributions, the ratio between the 
rates of bulge-bulge and disk-bulge events is $\Gamma_{\rm b}:
\Gamma_{\rm d} =61.8:38.2$.

We assign the lens mass based on the model mass function of \cite{gould00}.  
The model mass function is composed of stars, brown dwarfs (BDs), and 
stellar remnants of white dwarfs (WDs), neutron stars (NSs), and black 
holes (BHs).  The model is constructed under the assumption that bulge 
stars formed initially according to a double power-law distribution of 
the form
\begin{equation}
{dN\over dm} =
k \left( {m\over m_{\rm c}}\right)^\gamma;\qquad
\gamma =
\cases{
-2.0  & for $m>m_{\rm c}$, \cr
-1.3  & for $m<m_{\rm c}$,\cr
} 
\label{eq5}
\end{equation}
where $m_{\rm c}=0.7\ M_\odot$.  These slopes are consistent with the 
observations of \citet{zoccali00} except that the profile is extended 
to a BD cutoff of $m=0.03\ M_\odot$.  Based on this initial mass function, 
remnants are modeled by assuming that the stars with initial masses 
$1\ M_\odot< m <8\ M_\odot$, $8\ M_\odot < m < 40\ M_\odot$, and 
$m>40\ M_\odot$ have evolved into WDs (with a mass $0.6\ M_\odot$), 
NSs (with a mass $13.5\ M_\odot$), and BHs (with a mass $5\ M_\odot$), 
respectively. Note that since we assume stars with $m>1\ M_\odot$ have 
evolved into remnants, the upper limit of the stellar lens mass is 
$1\ M_\odot$, corresponding to the turn-off star in the Galactic 
bulge.\footnote{We note that disk turn-off is a bit brighter.  This may 
affect our result because it is easier to detect planets in the habitable 
zone with microlensing for more massive primaries.  However, the effect 
will be small because (1) the minority of events are due to disk lenses, 
(2) the disk turn-off is not much brighter than the bulge one, and so 
(3) the number of potential main-sequence lenses with mass significantly 
higher than a solar mass is quite small.}  Then, the resulting mass
fractions of the individual lens components are ${\rm stars}:{\rm BD}:
{\rm WD}:{\rm NS} :{\rm BH}=62:7:22:6:3$.  An important fraction of 
lenses are stars and thus they will contribute to the apparent brightness 
of the source star.  We consider this by determining the lens brightness 
by using the mass-$M_V$ relation listed in \cite{allen00}.

For each event involved with a primary lens, we then introduce a 
companion of an Earth-mass planet.  The location of the planet around 
the primary star is allocated under the assumption of a circular orbit 
with a power-law distribution of semi-major axes $a$, i.e. 
$dN_{\rm p}/da \propto a^{-\alpha}$, and random orientation of the 
orbital plane. There is little consensus about the power of the 
semi-major axis distribution.  \citet{tabachnik02} claimed $\alpha=1$ 
from the analysis of observed extrasolar planets detected by radial 
velocity surveys. The minimum mass solar nebula has surface density 
with $\alpha=1.5$ \citep{hayashi95}.  \citet{kuchner04} argued that 
multi-planet extrasolar planetary systems indicate $\alpha=2.0$.  We, 
therefore, test three different powers of $\alpha=1.0$, 1.5, and 2.0.   
The range of the semi-major axis is $-2\leq \log(a/{\rm AU})\leq 2$,
which is wide enough to cover the lensing zone of all possible lenses, 
and thus the exact range we consider does not affect the relative 
frequencies we calculate here.  Once a planetary event is produced, 
the rate of each event is computed by
\begin{equation}
\Gamma_{{\rm p},i} \propto \rho(D_{\rm S})D_{\rm S}^2 \rho(D_{\rm L}) 
\sigma_L v \Delta r_{\rm c},
\label{eq6}
\end{equation}
where $\rho(D)$ is the matter density along the line of sight, the 
factor $D_{\rm S}^2$ is included to account for the increase of the 
number of source stars with the increase of $D_{\rm S}$, $\sigma_L$ 
represents the lensing cross-section corresponding to the diameter of 
the Einstein ring, i.e.\ $\sigma_L=2\re$, and $v$ is the transverse 
speed of the lens with respect to the source star.  The factor 
$\Delta r_{\rm c}$ is included to weight the planet detection efficiency 
by the cross-section of the planetary perturbation under the assumption 
that the planet detection efficiency is proportional to the caustic 
size.\footnote{According to the formalism in equation~(\ref{eq3}), the 
size of the caustic scales as $s^{-2}$ for wide-separation planets 
($s\gg 1$), whereas the detection probability actually scales as $s^{-1}$.  
Then, the assumption that the planet detection efficiency is proportional 
to the caustic size fails to apply to these wide-separation planets.  
However, we note that the habitable zone is, in most cases, much smaller 
than the Einstein ring radius, and thus this failure does not affect our 
result.}  For the acceleration of the computation, we use $\Delta r_{\rm c}$ 
computed analytically by using the formalism derived under the 
perturbative approximation (Eqs.~[\ref{eq3}] and [\ref{eq4}]) except 
the region where the approximation is not valid.  In this region, we 
use the mean value of the numerical results averaged over the region 
(the solid curve in Fig.~\ref{fig:one}).\footnote{The caustic size 
presented in Fig.~\ref{fig:one} is for an Earth-mass planet around a 
primary with a fixed mass of $0.3\ M_\odot$, while the events produced 
by the simulation are associated with primary stars of various masses.  
We note, however, that for a given mass of the planet, $m_{\rm p}$, 
the physical size of the caustic does not depend on the mass of the 
primary, $m_\star$, because $\Delta r_{\rm c}\sim \Delta r_{\rm pc} 
\propto q^{1/2} \re \propto (m_{\rm p}/m_\star)^{1/2} m_\star^{1/2} 
= m_{\rm p}$.} Then, the fraction of events with planets in the 
habitable zone among the total number of events with detectable 
Earth-mass planets is calculated by
\begin{equation}
f_{\rm HZ}=
{ \sum \Gamma_{{\rm p},i} (d_{\rm in}\leq a \leq d_{\rm out})
\over 
\sum \Gamma_{{\rm p},i} },
\label{eq7}
\end{equation}
where $d_{\rm in}$ and $d_{\rm out}$ are the inner and outer limits 
of the habitable zone, respectively.  The semi-major axis is related 
to the projected separation by
\begin{equation}
a={s \over (\cos^2\phi + \sin^2\phi \cos^2 i)^{1/2}} \re,
\label{eq8}
\end{equation}
where $\phi$ is the phase angle of the planet measured from the 
major axis of the apparent orbit of the planet around the host 
star and $i$ is the inclination angle of the orbital plane.

% Figure 2 --------------------------------------------------------------
\begin{figure}[t]
\epsscale{1.2}
\plotone{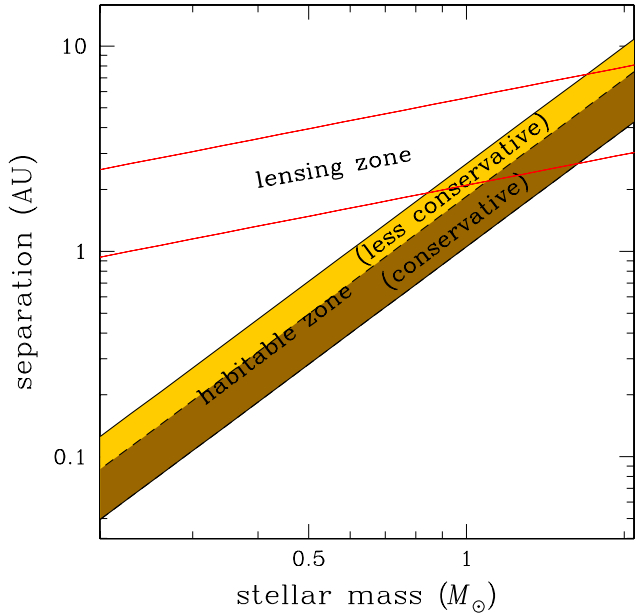}
\caption{\label{fig:two}
The ranges of the habitable zone and lensing zone as a function of 
a stellar mass.  For the habitable zone, the dark and light-shaded 
regions represent the conservative and less-conservative ranges of 
the habitable zone, respectively.  The lensing zone is computed 
assuming that the distances to the lens and source star are $D_L=6$ 
kpc and $D_S=8$ kpc, respectively.
}\end{figure}

The habitable zone is defined as a shell region around a star within 
which a planet may contain liquid water on its surface.  A convenient 
way of estimating the inner and outer limits of the habitable zone is 
to assume that the planet behaves as a graybody with an albedo and 
with perfect heat conductivity (implying that the temperature is 
uniform over the planet's surface).  Under this assumption, the 
inner and outer limits of the habitable zone are determined as
\begin{equation}
d_{\rm in(out)} = 
\left[ {(1-{\cal A}_L)L_\star \over 16\pi \sigma T^4} \right]^{1/2},
\label{eq9}
\end{equation}
where $L_\star$ is the luminosity of the star, ${\cal A}_L$ is the 
albedo of the planet, $\sigma$ is the Stefan-Boltzmann constant, and 
$T$ is the radiative equilibrium temperature.  In the simulation, we 
test two ranges of the habitable zone.  For a {\it conservative} range, 
we adopt radiative equilibrium temperatures of $T=269$ K and $203$ K 
for the inner and outer limits of the habitable zone, respectively.  
A {\it less-conservative} range has the same inner radius as that of 
the conservative range, but the outer limit is defined by a lower 
radiative equilibrium temperature of $T=169$ K \citep{lopez05}.  For 
the albedo, we choose ${\cal A}_L=0.2$ as a representative value of 
an Earth-like planet.  For a sun-like star, then, the inner limit of 
the habitable zone is $d_{\rm in}=0.96\ {\rm AU}$ and the outer limits 
are $d_{\rm out}=1.69\ {\rm AU}$ and 2.43 AU for the conservative and 
less-conservative ranges, respectively.  For the conversion of the 
mass of the star into the luminosity, we use the mass-luminosity 
relation \citep{allen00} of 
\begin{equation}
\log\left( {L\over L_\odot}\right) = 
3.8 \log \left( {m\over M_\odot}\right) + 0.08.
\label{eq10}
\end{equation}
Figure~\ref{fig:two} shows the conservative (dark-shaded region) and 
less-conservative (light-shaded region) ranges of the habitable zone 
as a function of the stellar mass.

The habitable zone is defined only for planets associated with 
stellar population of lenses.  Therefore, identifying the lens as a 
star is important to further isolate the sample of candidate planets 
in the habitable zone.  Fortunately, the proposed space-based lensing 
survey can sort out events associated with stellar lenses (hereafter 
bright-lens events) by analyzing the flux from the lens star 
\citep{bennett02, han05b}.  We, therefore, estimate an additional 
fraction of planets in the habitable zone among the planets involved 
with bright-lens events, $f_{\rm HZ,\star}$.  We assume that a lens 
can be identified as a star if it contributes $>10\%$ of the total 
observed flux.  Under this condition, we find that bright-lens events 
account for $\sim 22\%$ of the total events ($\sim 17\%$ of bulge-bulge 
and $\sim 29\%$ of disk-bulge events).  Identifying the stellar nature 
of the lens is also important to determine the distance to the lens 
and mass of the lens.  Once $D_L$ and $m$ are known, the projected 
star-planet separation in physical units is determined by $\tilde{d}=s\re$.

% Figure 3 --------------------------------------------------------------
\begin{figure}[t]
\epsscale{1.2}
\plotone{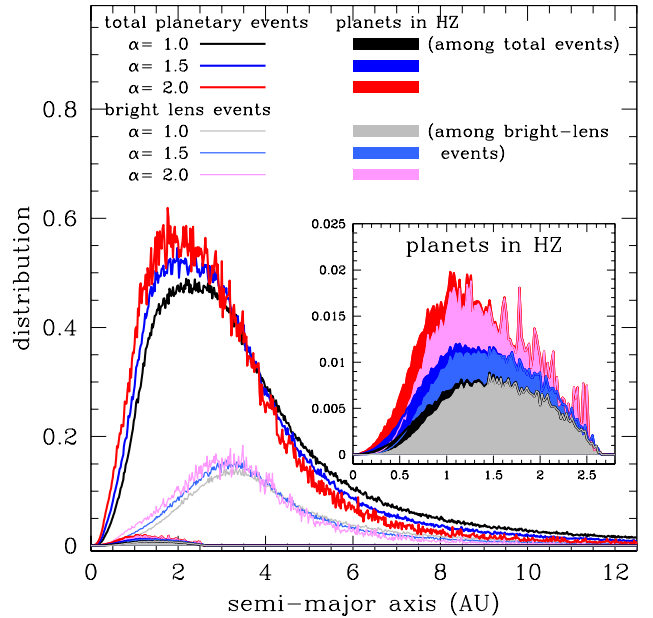}
\caption{\label{fig:three}
The distributions of the semi-major axes of Earth-mass planets to be 
detected in future lensing surveys.  The intrinsic semi-major axes of 
planets are assumed to be distributed by a power-law of the form 
$dN_{\rm p}/da \propto a^{-\alpha}$.  The thick curves with black (for 
$\alpha=1.0$), blue ($\alpha=1.5$), and red ($\alpha=1.0$) colors 
represent the distributions of all planets, while the curves with grey 
($\alpha=1.0$), light-blue ($\alpha=1.5$), and pink ($\alpha=2.0$) 
colors are the distributions of planets involved with bright lenses. 
The curves shaded with the corresponding colors are the distributions 
of planets located in the habitable zone.  
}
\end{figure}

% Table 1 ---------------------------------------------------------------
\begin{deluxetable*}{clcc}
\tablecaption{Fraction of Planets in the Habitable Zone \label{table:one}}
\tablewidth{0pt}
\tablehead{
\multicolumn{1}{c}{distribution} &
\multicolumn{1}{c}{definition of the} &
\multicolumn{2}{c}{planet fraction in the habitable zone} \\
\multicolumn{1}{c}{model} &
\multicolumn{1}{c}{habitable zone} &
\multicolumn{1}{c}{out of total events} &
\multicolumn{1}{c}{out of bright-lens events}}
\startdata
$dN_{\rm p}/da \propto a^{-1.0}$  & conservative       & 0.16\% & 0.71\% \\ 
\smallskip
                  -               & less-conservative  & 0.62\% & 2.70\% \\

$dN_{\rm p}/da \propto a^{-1.5}$  & conservative       & 0.26\% & 1.15\% \\ 
\smallskip
                  -               & less-conservative  & 0.88\% & 3.80\% \\

$dN_{\rm p}/da \propto a^{-2.0}$  & conservative       & 0.42\% & 1.74\% \\ 
                  -               & less-conservative  & 1.29\% & 5.40\% \\ 
\enddata
\tablecomments{ 
The fraction of habitable-zone Earth-mass planets to be detected in future 
lensing surveys.  We present two fractions.  One is the fraction out of 
the total number of planets ($f_{\rm HZ}$) and the other is the fraction 
among the planets involved with bright-lens events for which the primary 
lenses can be identified as stars ($f_{\rm HZ,\star}$).  For the definitions 
of the conservative and less-conservative ranges of the habitable zone, see 
\S\ 3.  The models in the first column indicate the assumed distributions of  
the intrinsic semi-major axis of planets.
}
\end{deluxetable*}

\section{Results and Discussion}

In Table~\ref{table:one}, we summarize the resulting fractions of 
events with planets located in the habitable zone determined from 
the simulation.  In Figure~\ref{fig:three}, we also present the  
semi-major axis distributions of the Earth-mass planets to be detected 
by the future lensing survey.  In the figure, the thick curves with 
black (for the power $\alpha=1.0$ of the intrinsic semi-major axis 
distribution), blue ($\alpha=1.5$), and red ($\alpha=2.0$) colors 
represent the distributions of all planets, while the curves with 
grey ($\alpha=1.0$), light-blue ($\alpha=1.5$), and pink ($\alpha=
2.0$) colors are the distributions for planets involved with bright 
lenses.  The curves shaded with the corresponding colors are the 
distributions of planets located in the habitable zone.  Here we use
the less-conservative range.

From the table and figure, we find the following trends.
\begin{enumerate}
\item
Planets located in the habitable zone comprises a very minor fraction 
of all planets.  We find that the fraction ranges $f_{\rm HZ}\sim 0.2\%$
-- 1.3\% depending on the model distributions of the intrinsic semi-major 
axes and 
the definitions of the habitable zone.  \citet{bennett02} predicted 
that the total number of Earth-mass planets that can be detected from 
the proposed space-based mission is $\sim 200$ if every lens star has 
an Earth-mass planet at an optimal position of 2.5 AU.  Even under 
this optimistic expectation, then, the number of planets in the habitable 
zone will be at most $\sim 3$, which is not adequate enough for the 
statistical analysis on the frequency of terrestrial planets in the 
habitable zone.
\item
Planets are biased toward smaller semi-major axis with the increase of 
the power $\alpha$ of the intrinsic semi-major axis distribution (c.f.\ 
the curves drawn with black [$\alpha=1.0$], blue [$\alpha=1.5$], and 
red [$\alpha=2.0$] colors).  However, the dependence of the resulting 
semi-major axis distributions of lensing planets on $\alpha$ is weak.  
\item
Planets of bright lenses are biased toward larger semi-major axis 
compared to the distribution of the total planets. This is because bright 
lenses tend to be heavier and thus have larger mean radius of the lensing 
zone, i.e.\ $r_{\rm E}$.  
\item
Most of habitable-zone planets are associated with bright lenses (c.f.\ 
the pairs of curves shaded with black-gray, blue-light blue, and 
red-pink colors). This is because the gap between the lensing zone and 
habitable zone becomes narrower with the increase of the lens brightness.  
If the sample is confined only to planets involved with bright lens 
events, as a result, the fraction of habitable-zone planets increases 
into $f_{\rm HZ,\star}\sim 0.7\%$ -- 5.4\%.
\end{enumerate}

Then, why is the microlensing sensitivity to Earth-mass planets located 
in the habitable zone so low?  First of all, the projected star-planet 
separation at which the planet detection sensitivity becomes maximum is 
in most cases substantially larger than the median value of the habitable 
zone.   This can be seen in Figure~\ref{fig:two} where the ranges of the 
lensing zone and habitable zone are plotted as a function of the lens mass.  
Roughly, the median radius of the habitable zone is linearly proportional 
to the lens mass, while the mean value of the lensing zone is proportional 
to the square root of the mass.  We find that the ratio of the median 
radius of the habitable zone to the mean radius of the lensing zone is 
roughly expressed as 
\begin{equation}
{d_{\rm HZ}\over r_{\rm E}} \sim
0.2 \left( {m\over 0.5\ M_\odot} \right)^{1/2}.
\label{eq11}
\end{equation}
As a result, the two zones overlap only for stars more massive than the 
sun \citep{distefano99}.  In addition, the habitable zone is defined by 
the {\it intrinsic} separation, while the lensing zone is defined by the 
{\it projected} separation.  Considering the projection effect, then, 
the gap between the lensing zone and habitable zone further increases.

\section{Conclusion}
We investigated the microlensing sensitivity of future lensing surveys 
to Earth-mass planets located in the habitable zone by conducting 
detailed simulation of Galactic microlensing events.  From the 
investigation, we found that the planets located in the habitable zone 
would comprise a very minor fraction and thus statistical analysis on 
the frequency of terrestrial planets in the habitable zone based on 
the sample would be very difficult.  We find the main reason for the 
low sensitivity is that the projected star-planet separation at 
which the microlensing planet detection efficiency becomes maximum 
is in most cases substantially larger than the median value of the 
habitable zone.

\acknowledgments 
We would like to thank B.\ S.\ Gaudi and A.\ Gould for making helpful
comments.
B.-G.\ P., Y.-B.\ J., \& C.-U.\ L.\ acknowledge support from a grant 
of the Korea Astronomy and Space Science Institute (KASI).  The work 
by C.H.\ was supported by the Astrophysical Research Center for the 
Structure and Evolution of the Cosmos (ARCSEC) of Korea Science and 
Engineering Foundation (KOSEF) through Science Research Program (SRC) 
program.

\end{document}